\documentclass[amsmath,amssymb,11pt]{revtex4}
\usepackage{amsfonts}
\usepackage{float}
\usepackage[colorlinks,linkcolor=blue,citecolor=green]{hyperref}
\usepackage{graphicx,color}
\usepackage{bm}

\begin{document}

\title{Generalized Ensemble Theory with Non-extensive Statistics}

\author{Ke-Ming Shen}
\email{shenkm@mails.ccnu.edu.cn}
\affiliation{Key Laboratory of Quark $\&$ Lepton Physics (MOE) and Institute of Particle Physics,
Central China Normal University, Wuhan 430079, China}

\author{Ben-Wei Zhang}
\email{bwzhang@mail.ccnu.edu.cn}
\affiliation{Key Laboratory of Quark $\&$ Lepton Physics (MOE) and Institute of Particle Physics,
Central China Normal University, Wuhan 430079, China}

\author{En-Ke Wang}
\affiliation{Key Laboratory of Quark $\&$ Lepton Physics (MOE) and Institute of Particle Physics,
Central China Normal University, Wuhan 430079, China}

\date{\today}

\begin{abstract}
 The non-extensive canonical ensemble theory is reconsidered with the method of Lagrange multipliers by maximizing Tsallis entropy, with the constraint that the normalized term of Tsallis' $q-$average
of physical quantities, the sum $\sum p_j^q$, is independent 
of the probability $p_i$ for Tsallis parameter $q$. The self-referential problem in the deduced probability and thermal quantities in non-extensive statistics is thus avoided, and thermodynamical relationships are obtained in a consistent and natural way. We also extend the study to the non-extensive
grand canonical ensemble theory and obtain the $q$-deformed Bose-Einstein distribution as well as the 
$q$-deformed Fermi-Dirac distribution. The theory is further applied to the generalized Planck law to 
demonstrate the distinct behaviors of the various generalized $q$-distribution functions discussed 
in literature.  

\end{abstract}

\maketitle

\section{Introduction}\label{sec:sec1}

In the last thirty years the non-extensive statistical mechanics,
based on Tsallis entropy~\cite{Tsallis-1988, Tsallis-book} and 
the corresponding deformed exponential function, has been developed and attracted a lot  of attentions with a large amount of applications in rather diversified fields~\cite{application}. 
Tsallis non-extensive statistical mechanics is a generalization of the common Boltzmann-Gibbs (BG) statistical mechanics by postulating a generalized entropy of the classical one,
$S=-k \sum _{i=1}^Wp_i \ln p_i$:
\begin{eqnarray}
S_q=-k \sum _{i=1}^Wp_i^q \ln_q p_i~,
\label{entropy-1}
\end{eqnarray}
where $k$ is a positive constant and denotes Boltzmann constant in BG statistical mechanics. For simplicity in the following we set $k=1$. Here the $q-$logarithm, $\ln_q x\equiv \frac{x^{1-q} -1}{1-q}$,
is introduced. 
In Eq.~(\ref{entropy-1})
$p_i$ stands for the probability distribution assigned to the microscopic configuration $i$,  $W$ gives the total number of the microscopic configuration in the system, and 
$q\in R$ is the so-called non-extensive parameter. One can check that it recovers
the BG statistics when $q\rightarrow 1$. 

With the generalized entropy at hand, the optimum probability distribution could be then obtained for this non-extensive entropy with the Maximum Entropy principle (MaxEnt), and a canonical ensemble theory of the non-extensive statistical mechanics  could be formulated. To do so, usually the method of Lagrange  multipliers (LM) will be used by including the constraint of energy conservation on the probability $p_i$. However, it is far from trivial to write down the total energy with the probability $p_i$ in non-extensive statistics and then have a well-defined partition function in canonical ensemble~\cite{Tsallis-book}.  Though progresses towards solving this problem have been made~\cite{Curado-1991,TMP-1998,Martinez-2000,Ferri-2005}, 
there are still some unpleasant confusions in the derivations, and further investigation and clarification should be needed to provide a consistent treatment.

Furthermore, as far as  the non-extensive quantum statistical mechanics is concerned, the generalized Bose-Einstein distribution for bosons  and  Fermi-Dirac distribution for fermions
(referred to as q-BED and q-FDD hereafter)
have been used~\cite{ Buyukkilic-1993, Buyukkilic-1995, Beck-2000, Teweldeberhan-2003, MPPP-2004,  Teweldeberhan-2005, Silva-2010, Conroy-2010}. 
However, in literature there is a confusion and debate on what is the proper expressions of q-BED and q-FDD.
In the work of \cite{Buyukkilic-1993, Buyukkilic-1995, Teweldeberhan-2003, Silva-2010, Conroy-2010, Teweldeberhan-2005},  the particle distribution functions 
in the non-extensive quantum statistics are shown as:
\begin{eqnarray}
\bar{n}_i=\frac{1}{e_{2-q}^{\alpha +\beta \epsilon _i}\pm 1}
\label{distribution-1}
\end{eqnarray}
(where $\alpha$ and $\beta$ are the Lagrange multiplier parameters, $\epsilon_i$ is
the energy of particle $i$, the upper sign is for fermions and lower one for bosons, respectively). 
Here the corresponding
deformed q-exponential function, $e_q^x:=[1+(1-q)x]^{1/(1-q)}$, is introduced. 
Whereas others~\cite{Beck-2000, MPPP-2004} have argued that the proper expressions of q-BED and q-FDD should be: 
\begin{eqnarray}
\bar{n}_i=\frac{1}{(e_{2-q}^{\alpha +\beta \epsilon _i})^q\pm 1} \,\,  ,
\label{distribution-2}
\end{eqnarray}
which is obtained by generalizing Hagedorn's statistical theory of momentum spectra of
particles produced in high-energy collisions, and has
also been widely used(for example, see~\cite{CTB-2010, De-2010}). We note that when $q \rightarrow 1$ both Eq.~(\ref{distribution-1})  and Eq.~(\ref{distribution-2}) 
go back to the conventional particle distributions in quantum statistics.
A comparison between these two expressions has been made by H. Hasegawa \cite{Hasegawa-2009}, where an analytical distribution with the $\Gamma(s)$ function and
the $(q-1)$ expansion of physical quantities have been provided.
And there are also different 
opinions \cite{Wilk-2008, Biro-2012} on whether it should have a
$q$-power index for the total number of particles, $N$, in the system. It is still not clear
how to derive the distribution functions of the non-extensive quantum statistical mechanics from first principles.

In this work we will revisit the canonical ensemble theory of the non-extensive statistics by imposing the constraint that $\frac{\partial}{\partial p_i}\sum _{j=1}^Wp_j^q= 0$ for any configuration $i$.  We show that the explicit self-referential problem will be avoided, and consistent thermodynamical relations could be obtained. The discussion is then extended to the non-extensive grand canonical ensemble and the corresponding q-BED and q-FDD are then derived, with which
an application to the generalized Planck law is also made to demonstrate the differences of several typical treatments of q-BED.

This paper is organized as follows. In Section~\ref{sec:sec2} we give a consistent derivation of the non-extensive canonical ensemble theory and obtain the corresponding thermodynamics relations. Next this study to the grand ensemble theory is extended in Section~\ref{sec:sec3}. Section~\ref{sec:sec4} gives a simple application of the developed model to the generalized Planck law in non-extensive statistical mechanics. A brief summary is then presented in Section~\ref{sec:sec5}.


\section{Canonical ensemble in non-extensive statistics}\label{sec:sec2}

\subsection{Tsallis $q-$Probability Distribution}

Firstly we study the classical case with the generalized entropy of Eq.(\ref{entropy-1})
in canonical ensemble theory. Consider a closed system which, however, can exchange heat with its surroundings and as a consequence will have a fluctuating total energy. For a closed equilibrium system,
the average energy is then fixed.  The $q-$average energy of the system
gives,
\begin{eqnarray}
U_q=\sum _i^W\frac{p_i^q}{\sum _{j=1}^W p_j^q}\epsilon _i
:=\sum _i^W P_i\epsilon _i
\label{energy-1}
\end{eqnarray}
with  the escort probability $P_i=p_i^q/\sum _{j=1}^W p_j^q$  the escort probability.
Within the generalized canonical ensemble theory,
under the constraints imposed by the $q-$average energy of the system\cite{Tsallis-1998},
as well as the normalized condition of probabilities, a $\Psi$ function is built up,
\begin{eqnarray}
\Psi(p_i):=S_q[p_i]-\alpha (\sum_{i=1}^W p_i-1) -\beta (\frac{\sum _i^W p_i^q\epsilon _i}{\sum _j^W p_j^q}-U_q) ~,
\label{olm-1}
\end{eqnarray}
where $S_q[p_i]$ is nothing but the Tsallis entropy of Eq.(\ref{entropy-1}), $\alpha$ and $\beta$ the
Lagrange multipliers (LM) respectively.

By extremizing entropy with MaxEnt,  when the partial derivatives of $\Psi$ equal $0$ we have 
\begin{flalign}
~~~~~~~~~~~~~~~~~~~~~~~~p_i&=\frac{1}{\bar{Z}^{(1)}_q}[1-(1-q)\frac{\beta}{\sum _{j=1}^Wp_j^q}(\epsilon _i-U_q)]^{\frac{1}{1-q}}& \nonumber \\	
&:=\frac{1}{\bar{Z}^{(1)}_q}\exp_q[-\frac{\beta}{\sum _{j=1}^Wp_j^q}(\epsilon _i-U_q)] \,\,  .
\label{probability-1-Tsallis}
\end{flalign}
Here the generalized partition function is defined as:
\begin{flalign}
~~~~~~~~~~~~~~~~~~~~~~~~\bar{Z}^{(1)}_q&=\sum [1-(1-q)\frac{\beta}{\sum _{j=1}
^Wp_j^q}(\epsilon _i-U_q)]^{\frac{1}{1-q}} & \nonumber \\
&:=\sum \exp_q[-\frac{\beta}{\sum _{j=1}^Wp_j^q}(\epsilon _i-U_q)]
\label{partition-1-Tsallis}
\end{flalign}
which normalizes the probability distributions.  The deformed $q-$exponential function,
$\exp_q(x):=[1+(1-q)x]^{1/(1-q)}$ is introduced,
whose inverse function, $q-$logarithm, is $\ln_q(x):=\frac{x^{1-q}-1}{1-q}$
with $x>0$. 
And the results agree with Tsallis'  results in~\cite{Tsallis-book}. 

However, one observes that $p_i$ distribution of Eq.~(\ref{probability-1-Tsallis})
is explicitly self-referential because the right-hand-side (RHS) of Eq.~(\ref{probability-1-Tsallis}) contains the term $\sum _{j=1}^Wp_j^q$.
To solve this problem, the OLM (Optimal Lagrange Multipliers) method has been
introduced \cite{Martinez-2000, Ferri-2005}, where the constraint of energy conservation (\ref{energy-1}) 
is rewritten as
\begin{equation}
\sum _{i=1}^W p_i'^q (\epsilon_i -U_q)=0  \,\,\, ,
\label{OLM-energy}
\end{equation}
and one obtains that
\begin{eqnarray}
\Psi'(p_i'):=S_q[p_i']-\alpha' (\sum_{i=1}^W p_i'-1) -\beta' (\sum _{i=1}^W p_i'^q (\epsilon_i -U_q))   \,\, .
\label{olm-psi}
\end{eqnarray}

Now the generalized probability distribution does not depend explicitly on itself,
\begin{equation}
p_i'=\frac{1}{\bar{Z}'_q}[1-(1-q)\beta'(\epsilon _i-U_q)]^{\frac{1}{1-q}} ~.
\label{probability-1-OLM}
\end{equation}

Comparing $p_i$ of Eq.~(\ref{probability-1-Tsallis}) and $p_i'$ of Eq.~(\ref{probability-1-OLM}), one can show that there is a relation
\begin{equation}
\beta'=\frac{\beta}{\sum _{j=1}^W p_i'^q} ~.
\label{OLM-b}
\end{equation}

A close examination of OLM method tells that the disturbing term, $\sum _{j=1}^W p_i^q$, still exists, and in deriving the thermodynamics relations with OLM, the parameter $\beta$ is used, which implies that the probability derived in Eq.~(\ref{probability-1-OLM}) is rather implicitly self-referential. Generally,
$\frac{\partial}{\partial p_i}\sum _{j=1}^Wp_j^q\neq 0$, i.e. the term of $\sum _{j=1}^Wp_j^q$ is $p_i-$dependent.   Therefore, the corresponding thermodynamics with $p_i$ of Eq.~(\ref{probability-1-Tsallis}) or   $p_i'$ of Eq.~(\ref{probability-1-OLM}) may be not self-consistent, and special attention 
should be made in doing calculations.

\subsection{$q-$Probability Distribution}

Given the non-extensive parameter $q$ itself, which describes the departure of system from the
ideal thermal equilibrium of BG case, it is reasonable to argue that the term of $\sum _{j=1}^Wp_j^q$
characterizes the non-extensive properties of the whole system  and
should be not connected to $p_i$ explicitly (the same as the term $\sum _{j=1}^Wp_j =1$ 
in BG case with $q \rightarrow 1$), namely,
\begin{eqnarray}
\frac{\partial}{\partial p_i}\sum _{j=1}^Wp_j^q=\frac{\partial}{\partial p_i}C_q= 0 ~,
\label{q-independent}
\end{eqnarray}
with $C_q$ a $p_i-$independent constant and $C_1=1$ as is well known.
Then the new $\Psi$ function should be, with one more constraint of $\sum p_i^q=C_q$,
\begin{eqnarray}
\Psi_0(p_i):=S_q[p_i]-\alpha_0 (\sum_{i=1}^W p_i-1) -\beta_0 (\frac{\sum _i^W p_i^q\epsilon _i}{\sum _j^W p_j^q}-U_q)-\gamma_0(\sum p_i^q-C_q) ~,
\label{olm-0}
\end{eqnarray}
where $\alpha_0$, $\beta_0$ and $\gamma_0$ are the new Lagrangian Multipliers.
With respect to $\partial \Psi_0/\partial p_i=0$, we can obtain
the generalized $q-$probability distribution function as 
\begin{flalign}
~~~~~~~~~~~~~~~~~~~~~~~~p_i&=\frac{1}{\bar{Z}_q}[1-(1-q)\frac{\beta^*}{\sum _{j=1}^Wp_j^q}(\epsilon _i-U_q)]^{\frac{1}{1-q}}& \nonumber \\	
&=\frac{1}{Z_q}[1-(1-q)\frac{\beta}{\sum_{j=1}^Wp_j^q}\epsilon_i]^{\frac{1}{1-q}} ~,
\label{probability-1}
\end{flalign}
with the generalized partition function given by
\begin{flalign}
~~~~~~~~~~~~~~~~~~~~~~~~\bar{Z}_q&=\sum [1-(1-q)\frac{\beta^*}{\sum _{j=1}
^Wp_j^q}(\epsilon _i-U_q)]^{\frac{1}{1-q}} & \nonumber \\
&:=\sum \exp_q[-\beta^{*'}(\epsilon _i-U_q)]
\label{partition-1}
\end{flalign}
and
\begin{eqnarray}
Z_q=\sum \exp_q(-\beta^{'}\epsilon _i)  \,\,  .
\label{partition-2}
\end{eqnarray}
It is worthy to mention that during the derivations of Eq.(\ref{probability-1}),
the first partial derivative goes like,
\begin{flalign}
~~~~~~~~~~~~~~~~~~~~~~~~~~~~~~~~~~~~\frac{\partial}{\partial p_i}S_q&=
\frac{\partial}{\partial p_i} (-\sum _{i=1}^Wp_i^q \ln_q p_i) & \nonumber \\	
&=-qp_i^{q-1}\ln_q p_i-p_i^q\frac{1}{p_i^q} \nonumber \\
&=-qp_i^{q-1}\ln_q p_i-1~.
\label{entropy-partial-1}
\end{flalign}
Here the non-extensive entropy, $S_q$, is recognized as a sum of non-extensive entropies of each 
part of system, $S_i:=-\ln_q p_i$, with a $q-$weighting on each probability.
Moreover, $\beta^*=\frac{\beta_0}{1-\gamma_0(1-q)}$ is the generalized Lagrangian Multiplier 
with relations such as
$\beta^{'} =\beta^{*'}/[1+(1-q)\beta^{*'} U_q]$, 
$\beta^{*'}=\beta^*/\sum p_j^q$ and $\beta'=\beta/\sum p_j^q$.
All of them are the same when $q=1$, and
$\gamma_0$ can be implicitly determined by the constraint $\sum p_j^q =C_q$ as well.
Worthy to mention that, 
we denote $\beta^*=\frac{\beta_0}{1-\gamma_0(1-q)}$
as the generalized Lagrangian Multiplier in a more consistent way in the following study.

We note that in non-extensive statistics, the $q$-expectation of quantities
characterizes the non-extensivity in the system~\cite{Tsallis-book,Biro-2013}.
The thermal average of an physical quantity $A$ in Tsallis statistics can be expressed by 
the following normalized $q$-expectation value consistently as,
\begin{eqnarray}
\langle A \rangle_q\equiv \frac{\sum _{i=1}^W p_i^q A_i}{\sum _{j=1}^Wp_j^q}
\equiv \sum _{i=1}^W P_i A_i ~.
\label{q-expectation}
\end{eqnarray}
Here $\{A_i\}$ are the corresponding eigenvalues for each microscopic configuration $i$ of the system.
$q$ is considered to be a parameter which describes the specific properties of the interactions of components of the system, or the deviation of the system from the classical equilibrium. And 
$P_i\equiv\frac{p_i^q}{\sum _{j=1}^Wp_j^q}$ is the escort probability, and normalized naturally. 
Note that $\sum p_j^q$ is a $p_i-$independent constant, whose deviation from unity in some way demonstrates the non-extensive properties of
the system departing from the BG equilibrium.


\subsection{$q-$Thermodynamical Relations}

In order to show the generalized thermodynamical properties for the system
with non-extensive $q-$probability distribution function of Eq.(\ref{probability-1}),
we have,
\begin{eqnarray}
\sum_{i=1}^{W}p_i^q=(\bar{Z}_q)^{1-q}  \,\,\,  .
\label{relation-1}
\end{eqnarray}
Substituting the above formula into the previous
Tsallis entropy in Eq.(\ref{entropy-1}), we get the generalized relation of entropy $S_q$ and
the $q-$partition function $\bar{Z}_q$,
\begin{eqnarray}
S_q=\frac{\sum_{i=1}^{W}p_i^q-1}{1-q}=\frac{(\bar{Z}_q)^{1-q}-1}{1-q}=\ln _q \bar{Z}_q  \,\, .
\label{relation-2}
\end{eqnarray}
Similarly we can get the properties of the other $q-$partition function, $Z_q$,
\begin{eqnarray}
(Z_q)^{1-q}=\sum_{i=1}^{W}p_i^q-(1-q)\beta U_q  \,\,   ,
\label{partition-3}
\end{eqnarray}
which indicates the relation between these two generalized partition functions as
\begin{eqnarray}
Z_q=\bar{Z}_q e_q^{-\beta'U_q}
\label{relation-3}
\end{eqnarray}
or
\begin{eqnarray}
\ln_q \frac{Z_q}{\bar{Z}_q}=-\beta'U_q   \,\,   .
\label{relation-4}
\end{eqnarray}

While for the $q-$average energy, $U_q$, we obtain
\begin{flalign}
~~~~~~~~~~~~~~~~~~~~~~~~~U_q&=\frac{\sum p_i^q \epsilon _i}{\sum p_j^q}& \nonumber \\
&=\frac{1}{\sum p_j^q}\frac{\sum (e_q^{-\beta\epsilon _i/(\sum p_j^q)}
)^q\epsilon _i}{Z_q^q}&\nonumber \\
&=-\frac{\partial }{\partial \beta} \ln_q Z_q   \,\,  .
\label{relation-5}
\end{flalign}
Similarly, the $q-$generalized force is given by
\begin{eqnarray}
Y_q=-\frac{1}{\beta}\frac{\partial}{\partial y}\ln_q Z_q \,\,  ,
\label{relation-6}
\end{eqnarray}
where $y$ is the corresponding $q-$generalized coordinate.

With the above relations we could derive that 
\begin{flalign}
~~~~~~~~~~~~~\beta (dU_q -Y_qdy)&=-\beta d(\frac{\partial}{\partial \beta}\ln_q Z_q)
+\frac{\partial}{\partial y}\ln_q Z_q dy&\nonumber \\
&=-\beta d(\frac{\partial}{\partial \beta}\ln_q Z_q)-\frac{\partial}{\partial\beta}\ln_q Z_q
d\beta &\nonumber \\
&~~~ +\frac{\partial}{\partial\beta}\ln_q Z_q
d\beta +\frac{\partial}{\partial y}\ln_q Z_q dy
& \nonumber \\
&=d(\ln_q Z_q-\beta \frac{\partial}{\partial\beta}
\ln_q Z_q)   \,\,    ,
\label{relation-7}
\end{flalign}
which tells us nothing but that $\beta =1/T$, comparing with
the $q$-thermodynamical relation $dS_q = \frac{1}{T}(dU_q-Y_qdy)$. And then
\begin{eqnarray}
S_q=\ln_q Z_q -\beta \frac{\partial}{\partial\beta}
\ln_q Z_q  \,\,  .
\label{relation-8}
\end{eqnarray}
By now, we know that in this generalized statistical mechanics, all
the thermodynamical relationships are being kept in the similar way as in BG statistical mechanics.


\section{Grand canonical ensemble in non-extensive statistics}\label{sec:sec3}

\subsection{$q$-Thermodynamics}

For the generalized grand canonical ensemble within non-extensive statistics, 
we consider a system being contacted with heat and particle baths. The open system with
a given temperature $T$, volume $V$, and chemical potential $\mu$ in the non-extensive statistics could be treated in parallel with what we have done for the generalized canonical ensemble in Section~\ref{sec:sec2}.

Therefore, in the non-extensive grand ensemble the generalized non-extensive probability $p_s$ 
to each distinct microstate $s$ is given by the following $q$-exponential function
\begin{eqnarray}
p_s=\frac{1}{\Xi _q}e_q^{-\alpha' N-\beta' E_s}
\label{probability-2}
\end{eqnarray}
where $\alpha'\equiv \alpha/(\sum_{t} p_t^q)$ and $\beta'\equiv \beta/(\sum_t p_t^q)$
with $\alpha$ and $\beta$ the corresponding Lagrange parameters.
And $\Xi_q =\sum_N \sum_s e_q^{-\alpha' N-\beta' E_s}$ is 
the generalized $q$-deformed grand canonical partition function.

In the following we list the formulas to get the generalized quantities from
the generalized $q$-equilibrium probability distribution and the corresponding
grand canonical partition function.

\begin{enumerate}
    \item The $q$-average of number of particles:
\begin{eqnarray}
\bar{N}_q=\frac{\sum p_s^q N}{\sum p_t^q}
=\frac{1}{\sum p_t^q}\frac{1}{\Xi_q ^q}\sum_N \sum_s 
(e_q^{-\alpha' N-\beta' E_s})^qN=-\frac{\partial}{\partial \alpha}\ln_q \Xi_q
\label{number-relation-1}
\end{eqnarray}
	\item The $q$-average of the total energy:
\begin{eqnarray}
U_q=\frac{\sum p_s^q E_s}{\sum p_t^q}
=-\frac{\partial}{\partial \beta}\ln_q \Xi_q
\label{energy-relation-2}
\end{eqnarray}

	\item The $q$-generalized force:
\begin{eqnarray}
Y_q=-\frac{1}{\beta}\frac{\partial}{\partial y}\ln_q \Xi_q
\label{force-relation-2}
\end{eqnarray}

    \item Similar to the case in the canonical ensemble, in the grand canonical ensemble we have $dS_q=\frac{1}{T}
(dU_q-Y_qdy-\mu d\bar{N}_q)$,
\begin{eqnarray}
\beta =1/T, ~~~\alpha =-\mu /T   \,\, .
\label{thermodynamics-1}
\end{eqnarray}

\end{enumerate}

\subsection{q-BED and q-FDD}

Next we  consider the non-extensive quantum statistics to obtain the generalized q-BED and q-FDD.
In a system composed of nearly independent particles, for simplicity we assume that there is only one kind of particle, whose number in each energy level $l$ ($l=1,2,3,\cdots$) is $n_l$ with the corresponding energy $\epsilon_l$.
Note that even in the non-extensive quantum system, the total number of particles should be
still extensive as well as the total energy in every energy level, namely,
\begin{eqnarray}
N=\sum _l n_l,~~~E=\sum _l \varepsilon _l n_l
\label{number-1}
\end{eqnarray} 
which leads to it that
\begin{flalign}
~~~~~~~~~~~~~~~\Xi_q &=\sum_N \sum_s e_q^{-\alpha' N-\beta' E_s}
=\sum\limits_{\{n_l\}}e_q^{-\sum _l(\alpha' +\beta'\varepsilon _l)n_l}&\nonumber \\
&=\sum\limits_{\{n_l\}}
\prod ^{q}\limits_{l}e_q^{-(\alpha' +\beta'\varepsilon _l)n_l} =\prod ^{q}\limits_{l}\sum\limits_{n_l}
e_q^{-(\alpha' +\beta'\varepsilon _l)n_l}&\nonumber \\
&=\prod ^{q}\limits_{l} Z_q(l)
\label{partition-4}
\end{flalign}
where $\prod ^{q}\limits_{l}x_l=x_1 \otimes _q x_2 \otimes _q x_3  \cdots$
and $Z_q(l)=\sum\limits_{n_l}
e_q^{-(\alpha' +\beta'\varepsilon _l)n_l}$ is the $l_{th}-$partition function.
Meanwhile, some properties with the $q$-algebra are listed here for reference:
\begin{eqnarray}
x \otimes _q y\equiv (x^{1-q}+y^{1-q}-1)_+^{\frac{1}{1-q}}
=e_q^{\ln_q x+\ln_q y}  \,\, ,
\label{q-algebra-1}
\end{eqnarray}
\begin{eqnarray}
e_q^x \otimes _q e_q^y=e_q^{x+y}  \,\, ,
\label{q-algebra-2}
\end{eqnarray}
\begin{eqnarray}
\ln_q x+\ln_q y=\ln_q (x \otimes _q y)  \,\, ,
\label{q-algebra-3}
\end{eqnarray}
\begin{eqnarray}
e_q^{-x}\cdot e_{q'}^x=1  \,\, ,
\label{q-algebra-4}
\end{eqnarray}
where $q'=2-q$. More can be seen in \cite{ST-2007, AMS-2013} and others.

As for the generalized average occupation number of particles in each energy level,
\begin{flalign}
~~~~~~~~~~~~~~~~~~~~~\bar{n}_l&=\frac{\sum p_s^q n_l}{\sum p_t^q} &\nonumber \\
&=\frac{1}{\sum p_t^q}\frac{1}{\Xi_q ^q}\sum_N \sum_s 
(e_q^{-\alpha' N-\beta' E_s})^qn_l&\nonumber \\
&=\frac{1}{\sum p_t^q}\frac{1}{\Xi_q ^q}
[\sum\limits_ {n_l}n_l(e_q^{-(\alpha' +\beta'\varepsilon _l)n_l})^q]
\prod ^{q}\limits_{m\neq l}Z_q(m)^q&\nonumber \\
&=\frac{1}{\sum p_t^q}\frac{1}{Z_q(l)^q}
\sum\limits_ {n_l}n_l(e_q^{-(\alpha' +\beta'\varepsilon _l)n_l})^q&\nonumber \\
&=-\frac{\partial}{\partial \alpha}\ln_q Z_q(l)  \,\, .
\label{number-2}
\end{flalign}

With the above relation, the generalized q-FDD and q-BED are easily obtained:
\begin{enumerate}
	\item For fermions, $n_l$ can only be $0$ or $1$ because of Pauli exclusion principle,
\begin{eqnarray}
Z_q(l)^{FD}=\sum\limits_{n_l=0}^{1}
e_q^{-(\alpha' +\beta'\varepsilon _l)n_l}
= 1+e_q^{-(\alpha' +\beta'\varepsilon _l)}   \,\, ,
\label{partition-5}
\end{eqnarray}
then,
\begin{eqnarray}
\bar{n}_l^{FD}=-\frac{\partial}{\partial \alpha}\ln_q Z_q(l)
=\frac{1}{\sum p_t^q}(\frac{1}{e_{q'}^{\alpha' +\beta'\varepsilon _l}+1})^q  \,\, .
\label{number-fermion-1}
\end{eqnarray}

	\item For bosons, there is no constraint for the values of any $n_l$,
\begin{eqnarray}
Z_q(l)^{BE}=\sum\limits_{n_l=0}^{\infty}
e_q^{-(\alpha' +\beta'\varepsilon _l)n_l}
\approx \frac{1}{1-e_q^{-(\alpha' +\beta'\varepsilon _l)}} \, ,
\label{partition-6}
\end{eqnarray}
namely, $\ln_q Z_q(l)=-\ln_q(1-e_q^{-(\alpha' +\beta'\varepsilon _l)})$, so
\begin{eqnarray}
\bar{n}_l^{BE}=-\frac{\partial}{\partial \alpha}\ln_q Z_q(l)
=\frac{1}{\sum p_t^q}(\frac{1}{e_{q'}^{\alpha' +\beta'\varepsilon _l}-1})^q \, .
\label{number-boson-1}
\end{eqnarray}

\end{enumerate}

Since the term $\sum_t p_t^q$ is considered to be a constant, we finally obtain the 
generalized q-FDD and q-BED as
\begin{eqnarray}
\bar{n}_l=(\frac{1}{e_{q'}^{\alpha' +\beta'\varepsilon _l}\pm 1})^q
\label{number-3}
\end{eqnarray}
where the '$+$' sign gives the result for fermions and '$-$' for bosons respectively. 

\subsection{Discussions}

We then discuss a few interesting properties of $q$-distribution functions of Eq.(\ref{number-3}).
In the following we use the Bose-Einstein case without loss of generality.
\begin{enumerate}
	\item In the limit of $q \rightarrow 1$, the terms of higher orders of $(1-q)$ could be neglected as an approximation,
\begin{eqnarray}
\bar{n}_{BE}(\omega)=\frac{1}{(e_{q'}^{\alpha' +\beta'\omega}- 1)^q}\approx \frac{1}{(e_{q'}^{\alpha' +\beta'\omega})^q-1}
\label{number-4}
\end{eqnarray}
which agrees with C.~Beck's result, cf. Eq.(\ref{distribution-2}) in~\cite{Beck-2000}.
On the other hand, when $(1-q)$ is not small, from Eq.~(\ref{number-1}) we have $\bar{n}_l=(\bar{n}'_l)^q$  at the same energy level $l$. Thus,
\begin{eqnarray}
\bar{n}'_{BE}(\omega)=\bar{n}_l^{1/q}=(\frac{1}{(e_{q'}^{\alpha' +\beta'\omega}- 1)^q})^{1/q}=\frac{1}{e_{q'}^{\alpha' +\beta'\omega}- 1}
\label{number-5}
\end{eqnarray}
which recovers the result in Eq.(\ref{distribution-1}) as discussed in Section~\ref{sec:sec1}.

	\item To see the differences more clearly, we plot all the Bose-Einstein distributions
	with energy dependence by setting $\mu=0$ (namely, $\alpha =0$) in Fig.~(\ref{int-f1}). It shows that: firstly, the difference
	between Eq.(\ref{distribution-1}) and the classical Bose-Einstein distribution is larger
	in the high energy range; the function of Eq.~(\ref{number-4}) has crossed with the classical
	one in the middle part; and our result of Eq.~(\ref{number-3}) drops faster than the other
	two generalized distributions; at last, all of them become the same when $q\to 1$.

\begin{figure}[H]
\vskip0.04\linewidth
\centerline{
\includegraphics[width = 0.8\linewidth]{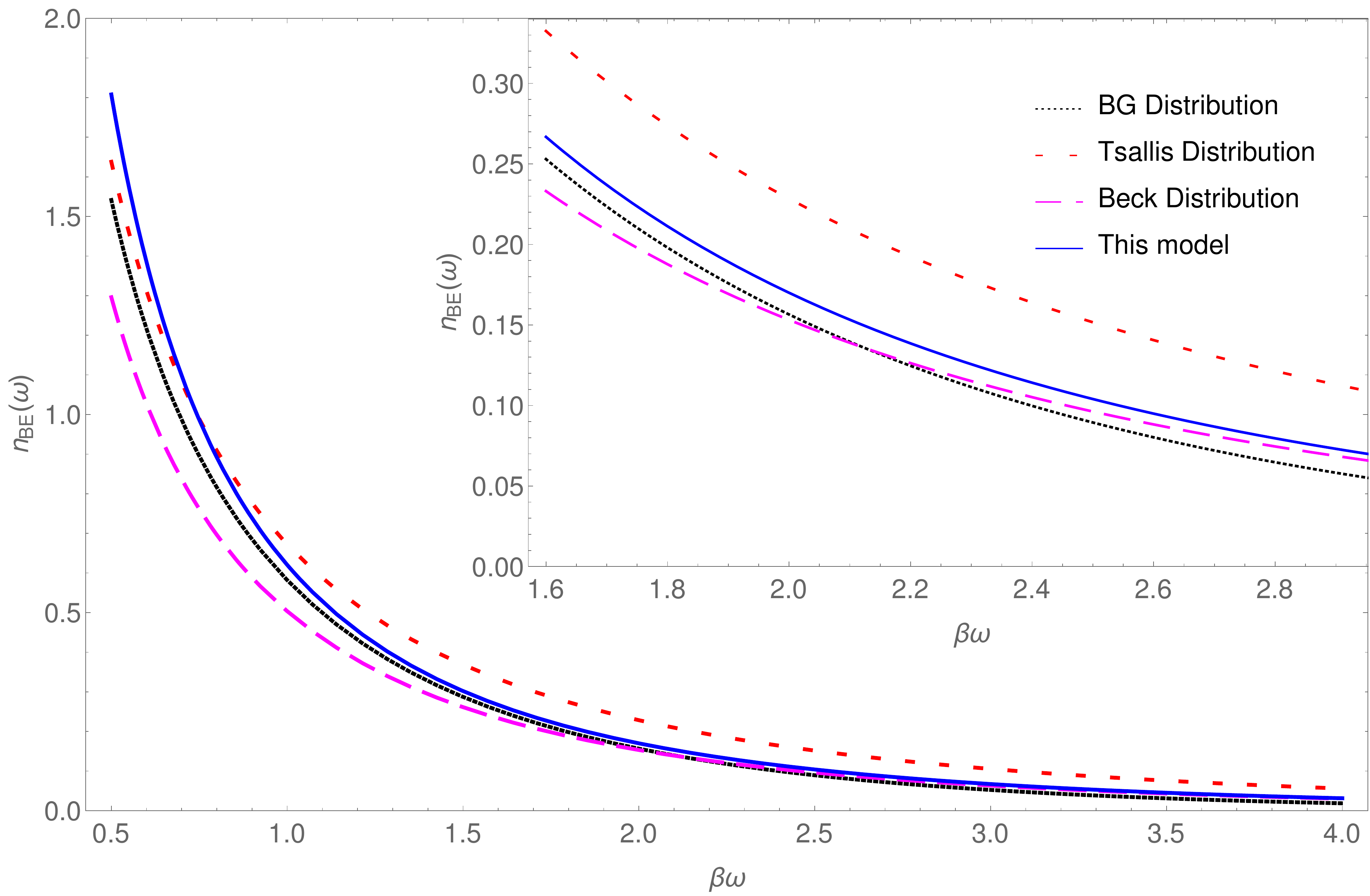}
}
\caption{  
Different generalized occupation number distribution functions for bosons,
with the chemical potential is set to zero and the Tsallis' parameter $q=1.2$. 
The black dotted curve, "BG", is 
the classical Bose-Einstein distribution; The red one with short dashed lines, "Tsallis", 
means the usual expression of Tsallis distribution, Eq.(\ref{distribution-1}); 
The purple long dashed one, "Beck", stands for the result
proposed by C. Beck of Eq.~(\ref{number-4}); and the blue curve gives the results with
the distribution function of Eq.~(\ref{number-3}) in our model.
}
\label{int-f1}
\end{figure}

	\item Moreover, note that the coefficient $\beta'$ is not the inverse function of temperature directly
	but with the extra constant term $\sum p_t^q$, its value depends on the Tsallis parameter $q$. 

	\item For all the above cases,the corresponding classical statistical results can be obtained
	naturally when $q\to 1$.

\end{enumerate}


\section{Application into the generalized Planck law}\label{sec:sec4}

\begin{figure}[H]
\vskip0.04\linewidth
\centerline{
\includegraphics[width = 0.5\linewidth]{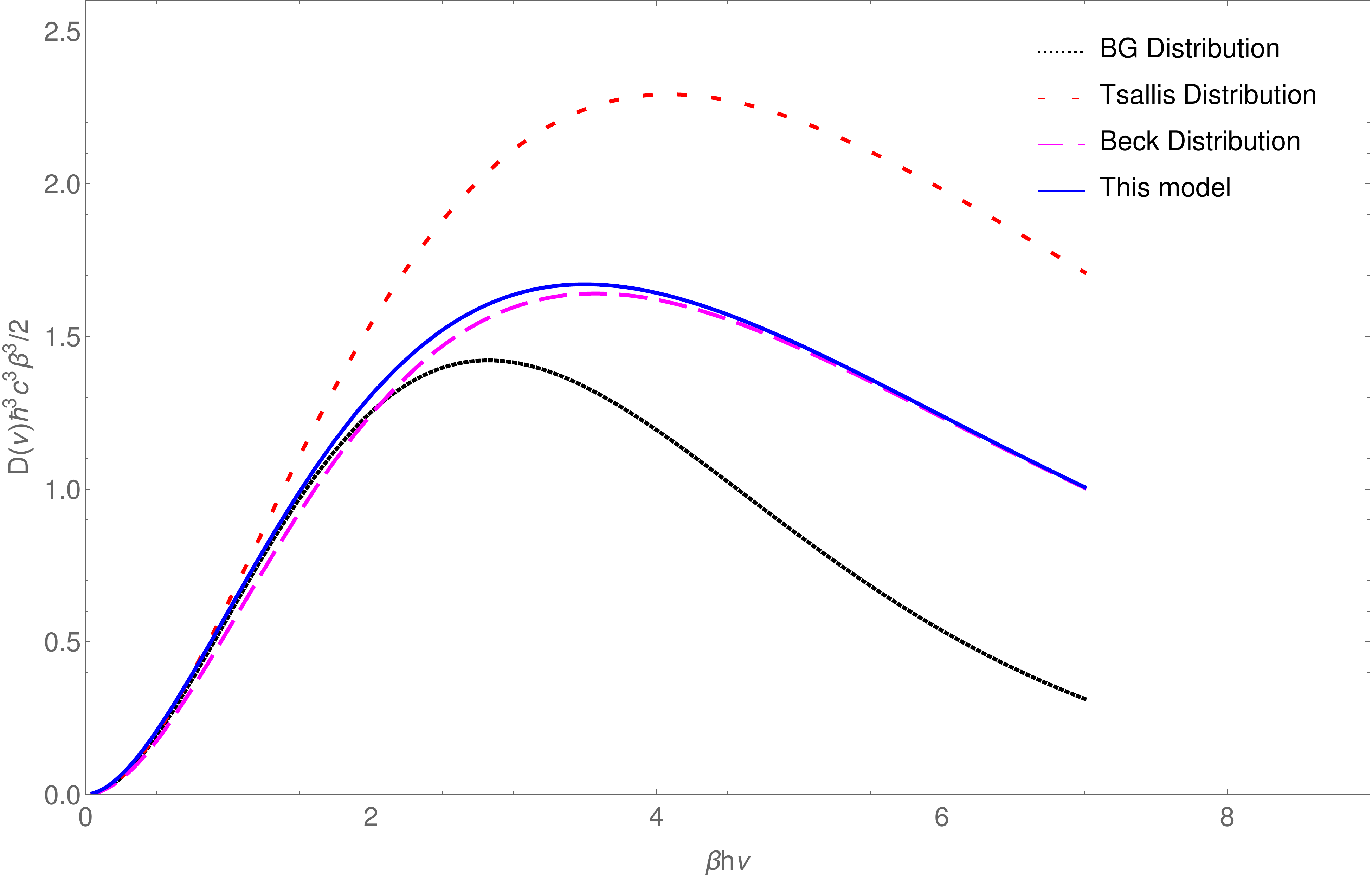}
\includegraphics[width = 0.5\linewidth]{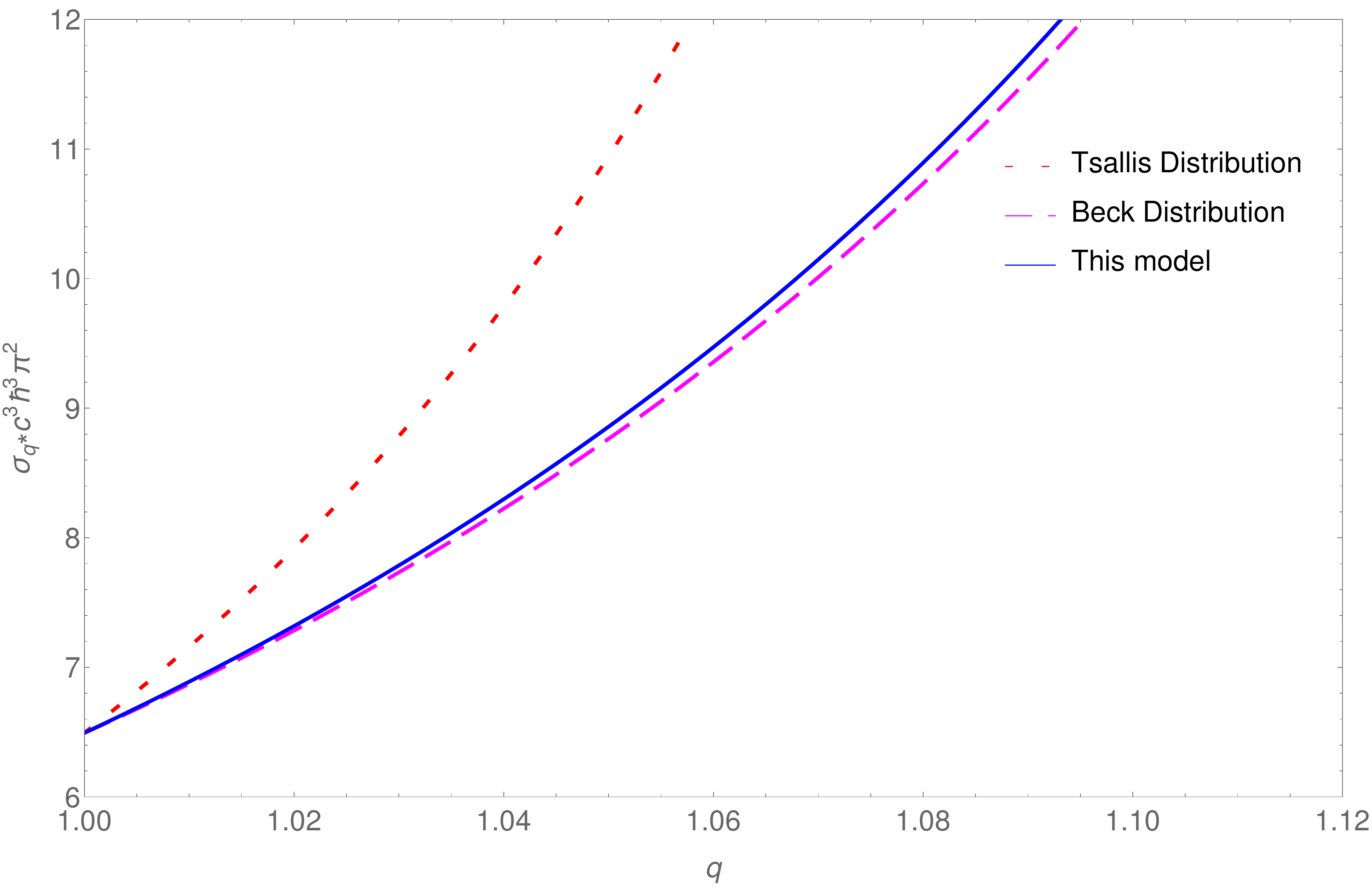}
}
\caption{  
{\bf Left panel}: Blackbody photon energy density per unit volume versus $\beta h\nu$
in the frame of Ref.~\cite{TBD-1998} of Eq.~(\ref{distribution-1}),
C. Beck's result of Eq.~(\ref{number-4}) and our results of Eq.~(\ref{number-3}) for $q=1.1$
as well as the classical case with $q=1.0$. 
{\bf Right panel}: The generalized Stefan-Boltzmann constant
from all different expressions as a function of the Tsallis' non-extensive parameter $q$,
where the starting point ($q=1$) is the classical result $\sigma\approx 6.49394$.
}
\label{int-f2}
\end{figure}

Using the results we have got, we rewrite the Planck law as 
\begin{eqnarray}
D_q(\nu)=\frac{8\pi T^3}{c^3h^2}\frac{x^3}{(e_{q'}^x-1)^q}
\label{planck-1}
\end{eqnarray}
where $x\equiv\beta h\nu$. And the comparison of the generalized Planck law 
within all the four different distributions is illustrated in the left panel of Fig.~(\ref{int-f2}).

Moreover, consider the total emitted power per unit surface,
\begin{eqnarray}
P_q=\int_0^{\infty} D_q(\nu)d\nu =\sigma _q T^4
\label{planck-2}
\end{eqnarray}
where $\sigma_q$ is the generalized Stefan-Boltzmann constant, and given by
\begin{eqnarray}
\sigma _q=\frac{8\pi}{c^3h^3}\int_0^{\infty}\frac{x^3}{(e_{q'}^x-1)^q}
\label{planck-3}
\end{eqnarray}
In the right panel of Fig.(\ref{int-f2}), we illustrate the generalized
Stefan-Boltzmann constant with three different $q$-distributions: Tsallis distribution 
in Eq.~(\ref{distribution-1})~\cite{TBD-1998}, Beck distribution in Eq.~(\ref{number-3})~\cite{Beck-2000}, and the distribution of our model in Eq.~(\ref{number-3}).  It is observed that the curve with our model has a similar trend as that with Beck distribution, and both curves goes up relatively much slower with $q$ as compared to the one of Tsallis distribution.


\section{Summary}\label{sec:sec5}

Starting with Tsallis entropy we have utilized the method of Lagrange multipliers to revisit the generalized canonical ensemble theory in non-extensive statistical mechanics, with the assumption that  $\sum_i p_i^q$ is independent of $p_i$, i.e., 
$\frac{\partial}{\partial p_i}\sum _{j=1}^Wp_j^q= 0 $. In our approach the (explicitly or implicitly) self-referential problem in the $q$-probability distribution and thermal quantities is bypassed, and a consistent theory of the non-extensive canonical ensemble is provided. The corresponding $q$-deformed thermodynamical relations have also been provided consistently, which naturally take similar forms as those in BG statistical mechanics.

Our results have also been extended to the generalized grand canonical ensemble case in the non-extensive quantum statistical mechanics. We have considered the non-extensive
effects on both the energy and the number of constituent particles of the system. The general distributions
of q-BED and q-FDD, as well as its thermodynamical properties have been obtained in the non-extensive quantum statistics.
Furthermore, we have also compared our results with other different expressions of $q$-distributions: Tsallis distribution and Beck distribution. We observe that by taking different limits of $q$, the results of our model may recover the formulas of $q$-distributions used by other authors.

Finally we have also included a simple application of our model by investigating the generalized Planck law with our
$q-$distributions of Eq.(\ref{number-3}), and the numerical simulations with  Tsallis distribution, Beck distribution as well as the distribution in our model have also been illustrated for clarification. It is found that the distribution of our model has similar behavior as Beck distribution, though appreciable difference between results with the distribution in our model and those by Tsallis distribution has been observed.

\vspace{0.6cm}
{\bf{Acknowledgements}}:
We'd like to thank T. S. Biro, G. Y. Ma, J. Yan and C. Zhang for helpful discussions. This work has been supported by NSFC of China with Project No. 11435004 and 11521064, and by MOST of China under 2014DFG02050.

\end{document}